\documentclass[pra,twocolumn,showpacs,letterpaper,showpacs,superscriptaddress]{revtex4}
\usepackage{graphicx,amsmath,amssymb,amsfonts,latexsym,color,dcolumn,bm,epsfig,subfigure}
\newcommand{\bd}{{\bf d}}

\newcommand{\eps}{\epsilon}

\newcommand{\be}{\begin{equation}}
\newcommand{\ee}{\end{equation}}
\newcommand{\bea}{\begin{eqnarray}}
\newcommand{\eea}{\end{eqnarray}}

\newcommand{\sig}{{\sigma}}

\newcommand{\ra}{\rangle}
\newcommand{\la}{\langle}
\newcommand{\om}{\omega}
\newcommand{\Om}{\Omega}

\newcommand{\bsub}{\begin{subequations}}
\newcommand{\esub}{\end{subequations}}
\newcommand{\baline}{\begin{eqalignno}}
\newcommand{\ealine}{\end{eqalignno}}

\begin{document}
\title{Velocity dependence of the quantum friction force on an atom near a dielectric surface} 
\author{P. W. Milonni}
\affiliation{Theoretical Division, MS B213, Los Alamos National Laboratory, Los Alamos, NM 87545, USA}
\affiliation{Department of Physics and Astronomy, University of Rochester, Rochester, NY 14627, USA} 

\date{today}

\begin{abstract}
Different approaches to the problem of the ``quantum friction" force $F$ acting on an atom moving with velocity $v\ll c$ parallel to a dielectric surface have resulted in different predictions for the way in which $F$ depends on $v$. For instance,
Scheel and Buhmann [Phys. Rev. A {\bf 80}, 042902 (2009)] and Barton [New J. Phys. {\bf 12}, 113045 (2010)] and others find a force
linear in $v$, while Intravaia, Behunin, and Dalvit [arXiv:1308.0712, 2013] and others find that the force varies as $v^3$ to lowest order in $v$. We argue that the $F\propto v^3$ prediction results from an oversimplified treatment of the atom as a linear oscillator, and that $F$ depends on $v$ in just the way predicted by Scheel and Buhmann and Barton.

\end{abstract}
\maketitle
Scheel and Buhmann \cite{sb2009} and Barton \cite{barton} have considered the force experienced by an atom moving with a velocity $v$ parallel to a dielectric surface. For unexcited two-level atoms (TLA) they derive by different methods the same resistive force with a linear dependence on $v$ \cite{buhmannbook}. Scheel and Buhmann express the force in terms of a dipole correlation function; the principal approximation made in the evaluation of this correlation function is essentially of the Weisskopf-Wigner type or, since it is made in the context of a two-time correlation function, what could be called a quantum regression ``theorem" (QRT).  Barton uses standard perturbation theory to obtain the same expression based on the kinetic energy lost by a moving two-level atom and transferred to plasmons of the dielectric. Scheel and Buhmann allow for multilevel atoms and atoms in excited states. While excited states will figure into the discussion below, the two-level model is adequate for our purposes.

Intravaia {\sl et al.} \cite{dalvit} find a force proportional to $v^3$. They claim that
the QRT fails for this system and also claim to base their calculation on the fluctuation-dissipation theorem (FDT) for a two-level atom. Regarding the FDT, we note that the version they invoke in their Eq. (2) is a standard result of {\sl linear} response theory
and does not apply to an atom (e.g., it carries no information as to the state of the atom). It should also be noted that in Barton's energy-conservation approach neither the FDT nor the QRT are needed. 

We have calculated the force on the TLA using the familiar expression for the force on a polarizable particle and working in the Heisenberg picture with a symmetric ordering of atom and field operators. We make the Weisskopf-Wigner approximation in which the upper state of the TLA decays exponentially at the rate $\gamma$, which is easy to calculate and varies as $1/z^3$, where $z$ is the (constant) distance of the atom from the surface. For times $t\gg 1/\gamma$, the contribution to the force from the free field (i.e., the part of the field that does not depend on the internal dynamics of the atom) is
\bea
F_0&\cong&-\frac{\hbar}{\pi^2}\la\sig_z(t)\ra\int d^2kk_xke^{-2kz}\nonumber\\
&&\mbox{}\times\int_0^{\infty}d\om\Delta_I(\om)\alpha_I(\om-k_xv),
\label{f0eq}
\eea
where $v$ is the (constant) velocity of the atom (along the $x$ direction), $\la\sig_z(t)\ra$ is the TLA population difference in the conventional notation,  
\be
\alpha_I(\om)=\frac{|\bd|^2}{3\hbar}\Big\{\frac{\gamma/2}{(\Om-\om)^2+\gamma^2/4}-\frac{\gamma/2}{(\Om+\om)^2+\gamma^2/4}
\label{alphaIeq}
\ee
is the imaginary part of the polarizability, $\bd$ is the transition electric dipole moment, and $\Delta_I(\om)$ is the imaginary
part of \cite{barton}
\be
\Delta(\om)=\frac{\eps(\om)-1}{\eps(\om)+1},
\label{deltadef}
\ee 
\be
\eps(\om)=1+\frac{\om_p^2}{\om_0^2-\om^2-i\om\Gamma} \ .
\label{heps}
\ee
For the contribution to the force from the source (reactive) part of the field we obtain similarly
\bea
&&F_S\cong-\frac{|\bd|^2}{3\pi^2}\int d^2kkk_xe^{-2kz}\int_0^{\infty} d\om\Delta_I(\om)\nonumber\\
&\times&\Big[\frac{\gamma/2}{(\Om+\om-k_xv)^2+\gamma^2/4}+\frac{\gamma/2}{(\Om-\om+k_xv)^2+\gamma^2/4}\Big].\nonumber\\
\eea
The complete force is
\begin{widetext}
\be
F=F_0+F_S\cong-\frac{2|\bd|^2}{3\pi^2}\int d^2kkk_xe^{-2kz}\int_0^{\infty}d\om\Delta_I(\om)
\Big[\frac{p_1\gamma/2}{(\Om+\om-k_xv)^2+\gamma^2/4}+
\frac{p_2\gamma/2}{(\Om-\om+k_xv)^2+\gamma^2/4}\Big],
\label{tla}
\ee
\end{widetext}
where $p_1=(1/2)(1-\la\sig_z\ra)$ and $p_2=(1/2)(1+\la\sig_z\ra)$ are respectively the lower- and upper-state TLA occupation
probabilities.

For a ground-state atom ($p_1=1,p_2=0)$, therefore, the friction force is
\bea
F&\cong&-\frac{|2\bd|^2}{3\pi^2}\int d^2kkk_xe^{-2kz}\int_0^{\infty}d\om\Delta_I(\om)\nonumber\\
&&\mbox{}\times\frac{\gamma/2}{(\Om+\om-k_xv)^2+\gamma^2/4},
\eea
which, to first order in $v$, agrees with the results of Scheel and Buhmann \cite{sb2009} and Barton \cite{barton}. For
an excited-state atom ($p_1=0,p_2=1$) we obtain similarly the result of Scheel and Buhmann.

Consider now a linear oscillator instead of a TLA. In this case $\sig_z(t)$ is replaced everywhere by $-1$, and
Eq. (\ref{f0eq}), for instance, is replaced by
\be
F_0=\frac{\hbar}{\pi^2}\int d^2kk_xke^{-2kz}
\int_0^{\infty}d\om\Delta_I(\om)\alpha_I(\om-k_xv).
\label{f0osceq}
\ee
In the oscillator case the Heisenberg equation of motion for $\sig_x(t)$ is {\sl linear} and the dipole correlation function required for the calculation of $F_S$ is given by the standard fluctuation-dissipation theorem \cite{landaulifshitz}; this results straightforwardly in the (zero-temperature) expression
\begin{widetext}
\be
F_S=\frac{\hbar}{\pi^2}\int_{-\infty}^{\infty}dk_y\int_0^{\infty}dk_xkk_xe^{-2kz}
\int_0^{\infty}d\om\Delta_I(\om)
\Big\{\alpha(\om+k_xv)-\alpha_I(\om-k_xv){\rm sign}(\om-k_xv)\Big\}.
\label{sho1}
\ee
\end{widetext}
After some algebra one finds for the linear oscillator
\bea
F&=&F_0+F_S=\frac{2\hbar}{\pi^2}\int_{-\infty}^{\infty}dk_y\int_{0}^{\infty}dk_xkk_xe^{-2kz}\nonumber\\
&&\mbox{}\times\int_0^{\infty}d\om\theta(k_xv-\om)\Delta_I(\om)\alpha_I(\om-k_xv),
\label{sho}
\eea
where $\theta$ is the unit step function. This equation is essentially the basis for the claim by Intravaia {\sl et al.} that the force varies as $v^3$ rather than $v$ to lowest order in $v$. It is not a new result, having been obtained some time ago by Kyasov and Dedkov \cite{dk2002}, although these authors have published several different and conflicting results. 

We conclude that Intravaia {\sl et al.} have not obtained a friction force for an atom, but rather a force
that follows from an oversimplified treatment of the atom as a linear oscillator. Since the linear oscillator model is very often an excellent approximation to an atom in an {\sl external} field, when the atom remains with high probability in its ground state, it may seem surprising that the model does not accurately describe the atom in the quantum friction problem. In fact it is well known that the linear oscillator model gives an incorrect expression even for the simple ``vacuum" frequency shift in the case of interest here in which the atom is near a dielectric surface \cite{agar}; in the oscillator model, but not in the TLA model, the polarizability appears in the integral over $\om$ for the shift. In the present problem, similarly, the imaginary part of the polarizability appears in the force (\ref{sho}) obtained with the oscillator model but not in the TLA force (\ref{tla}).

\end{document}